\documentclass[structabstract]{aa}  
\usepackage{graphicx}
\usepackage[varg]{txfonts}
\usepackage[round]{natbib}
\bibpunct{(}{)}{;}{a}{}{,} % to follow the A&A style
\usepackage{epsfig}
\usepackage{longtable}
\begin{document}

   \title{Radial velocity variations in the young eruptive star
     EX~Lup\thanks{This work is based in part on observations made
       with ESO Telescopes at the La Silla Paranal Observatory under
       program IDs 079.A-9017, 081.A-9005, 081.A-9023, 081.C-0779,
       082.C-0390, 082.C-0427, 083.A-9011, 083.A-9017, 084.A-9011,
       085.A-9027, 086.A-9006, 086.A-9012, 087.A-9013, 087.A-9029, and
       089.A-9007.}}

   \authorrunning{K\'osp\'al et al.}

   \titlerunning{RV variations in EX~Lup}

   \author{\'A. K\'osp\'al\inst{1}\thanks{ESA Fellow}
          \and
          M. Mohler-Fischer\inst{2}
          \and
          A. Sicilia-Aguilar\inst{3}
	  \and
          P. \'Abrah\'am\inst{4}
          \and
          M. Cur\'e\inst{5}
          \and
          Th. Henning\inst{2}
          \and
          Cs. Kiss\inst{4}
          \and
          R. Launhardt\inst{2}
          \and
          A. Mo\'or\inst{4}
          \and
          A. M\"uller\inst{6}
          }

\institute{Research and Scientific Support Department, European
Space Agency (ESA-ESTEC, SRE-SA), PO Box 299, 2200 AG, Noordwijk,
The Netherlands\\
\email{akospal@rssd.esa.int}
\and
Max-Planck-Institut f\"ur Astronomie, K\"onigstuhl 17,
69117 Heidelberg, Germany
\and
Departamento de F\'\i{}sica Te\'orica, Facultad de Ciencias,
Universidad Aut\'onoma de Madrid, 28049 Cantoblanco, Madrid, Spain
\and
Konkoly Observatory, Research Centre for Astronomy and
Earth Sciences, Hungarian Academy of Sciences, 
PO Box 67, 1525 Budapest, Hungary
\and
Departamento de F\'\i{}sica y Astronom\'\i{}a, Facultad de Ciencias,
Universidad de Valpara\'\i{}so, Av.~Gran Breta\~na 1111, Casilla 5030,
Valpara\'\i{}so, Chile
\and
European Southern Observatory, Alonso de Cordova 3107, Vitacura,
Santiago, Chile
}

\date{Received date; accepted date}

%-----------------------------------------------------------------
% ABSTRACT
%-----------------------------------------------------------------
 
  \abstract
  {EX~Lup-type objects (EXors) are low-mass pre-main sequence objects
    characterized by optical and near-infrared outbursts attributed to
    highly enhanced accretion from the circumstellar disk onto the
    star.}
  {The trigger mechanism of EXor outbursts is still debated. One type
    of theory requires a close (sub)stellar companion that perturbs
    the inner part of the disk and triggers the onset of the enhanced
    accretion. Here, we study the radial velocity (RV) variations of
    EX~Lup, the prototype of the EXor class, and test whether they
    can be related to a close companion.}
  {We conducted a five-year RV survey, collecting 54 observations with
    HARPS and FEROS. We analyzed the activity of EX~Lup by checking
    the bisector, the equivalent width of the Ca 8662\,\AA{} line, the
    asymmetry of the Ca\,II K line, the activity indicator S$_{\rm
      FEROS}$, the asymmetry of the cross-correlation function, the
    line depth ratio of the VI/FeI lines, and the TiO, CaH\,2, CaH\,3,
    CaOH, and H$\alpha$ indices. We complemented the RV measurements
    with a 14-day optical/infrared photometric monitoring to look
    for signatures of activity or varying accretion.}
 {We found that the RV of EX~Lup is periodic ($P=7.417\,$d), with
   stable period, semi-amplitude (2.2\,km\,s$^{-1}$), and phase over
   at least four years of observations. This period is not present in
   any of the above-mentioned activity indicators. However, the RV of
   narrow metallic emission lines suggest the same period, but with
   an anti-correlating phase. The observed absorption line RVs can be
   fitted with a Keplerian solution around a 0.6\,M$_{\odot}$ central
   star with M$_2\sin i$=(14.7$\pm$0.7)\,M$_{\rm Jup}$ and
   eccentricity of $e=0.24$. Alternatively, we attempted to model the
   observations with a cold or hot stellar spot as well. We
     found that in our simple model, the spot parameters needed to
     reproduce the RV semi-amplitude are in contradiction with the
     photometric variability, making the spot scenario unlikely.}
  {We qualitatively discuss two possibilities to explain the RV data:
    a geometry with two accretion columns rotating with the star, and
    a single accretion flow synchronized with the orbital motion of
    the hypothetical companion; the second scenario is more consistent
    with the observed properties of EX~Lup. In this scenario, the
    companion's mass would fall into the brown dwarf desert, which,
    together with the unusually small separation of 0.06\,AU would
    make EX~Lup a unique binary system. The companion also has
    interesting implications on the physical mechanisms responsible
    for triggering the outburst.}


   \maketitle

%-----------------------------------------------------------------
% INTRODUCTION
%-----------------------------------------------------------------

\section{Introduction}

EX~Lup-type objects (EXors) form a spectacular group of low-mass
pre-main sequence (PMS) objects, characterized by repetitive optical
outbursts of 1--5\,mag, lasting from a few months to a few years. The
outburst is usually attributed to enhanced accretion from the inner
circumstellar disk (within $\sim$0.1\,AU) to the stellar surface,
caused by an instability in the disk \citep{herbig1977,herbig2008}. In
quiescence, EXors typically accrete at a rate of 10$^{-10}$ to
10$^{-7}$\,M$_{\odot}$\,yr$^{-1}$, while in outburst, accretion rates
are about an order of magnitude higher \citep{lorenzetti2012a}. The
brief episodes of highly increased accretion may significantly
contribute to the build-up of the final stellar mass. Moreover, the
outbursts have a substantial effect on the circumstellar material. The
importance of the outburst phase on early stellar/disk evolution was
demonstrated by \citet{abraham2009}, for example, who discovered
episodic crystallization of silicate grains in the disk surface due to
the increased luminosity and temperature during the 2008 outburst of
\object{EX~Lup}, resulting in material that forms the building blocks
of primitive comets.

The origin of these accretion outbursts is still highly debated. One
type of explanation includes viscous-thermal instabilities in the disk
\citep{bell1994}, a combination of gravitational and
magneto-rotational instability \citep{armitage2001}, or accretion of
clumps in a gravitationally unstable disk
\citep{vorobyov2005,vorobyov2006}. Another model proposes that
accretion onto a strongly magnetic protostar is inherently episodic if
the disk is truncated close to the corotation radius
\citep{dangelo2010}. Yet another type of theory involves a close
stellar or sub-stellar companion that perturbs the disk and triggers
the onset of the enhanced accretion. The actual physical process could
be thermal instability induced by density perturbations due to a
massive planet in the disk \citep{lodato2004}, for example, or tidal
effects from close companions \citep{bonnell1992}. Such explanations
are especially favored in cases where the accretion rate increases by
several orders of magnitude in just a few weeks or few months, because
these rapid rise times are difficult to explain without external
perturbations.

It is an open question whether all young eruptive stars have
companions. Among the FU\,Orionis-type objects, which are young stars
producing even more powerful and longer lasting outbursts than EXors,
binarity is not unheard of. For instance, a companion to the prototype
object FU\,Ori was found with the direct imaging technique at a
projected distance of 225\,AU \citep{wang2004}. Some EXors are also in
binaries, such as the spectroscopic binary system UZ~Tau~E
\citep{jensen2007}, or the visual binaries V1118~Ori
\citep[separation 0$\farcs$18;][]{reipurth2007b} and VY~Tau
\citep[separation 0$\farcs$66;][]{leinert1993}. Since the triggering
mechanism requires a companion that perturbs the inner part of the
disk, typically at a few tenths of an AU, radial velocity (RV) methods
could be suitable to find such companions. However, many of the young
eruptive stars have never been searched for close companions with RV
methods, mostly because of the difficulty of measuring the RV in
young, chromospherically active and/or actively accreting stars.

In this paper we present a spectroscopic and photometric monitoring of
EX~Lup, the prototypical EXor object, focusing on the possible
existence of a close companion and on the accretion process. EX~Lup is
an M0.5-type young star situated not far from the Lupus 3 star-forming
region, at a distance of 155\,pc \citep{lombardi2008}. Over the last
six decades, EX~Lup has exhibited two large eruptions ($\Delta
V\,{\gtrsim}\,$5\,mag), and several smaller ($\Delta
V\,{=}\,$1-3\,mag) outbursts. The latest outburst happened in
2008. Since then, the system has been mostly quiescent, with minor
fluctuations around $V$=12-13\,mag. Several authors have searched for
a companion to EX~Lup with different methods without any
success. \citet{ghez1997} used $K$-band direct imaging and were able
to exclude the presence of a companion between about 150\,AU and
1800\,AU. \citet{bailey1998} used spectro-astrometry, but found no
companion down to 15\,AU. A few sporadic RV measurements with
different instruments (two values in \citealt{melo2003}, three values
in \citealt{guenther2007}, and four values in \citealt{herbig2007})
were inconclusive in terms of a companion.

This study is the continuation of a series of papers by our group
investigating the quiescent state and the extreme outburst of EX~Lup
in 2008 \citep{sipos2009, abraham2009, goto2011, kospal2011a,
  juhasz2012, sicilia2012}. In Sect.~\ref{sec:obs} we summarize the
observations and describe the steps of data reduction. In
Sect.~\ref{sec:res} we analyze the RV data and different stellar
activity indicators, present our spot model, and study the photometric
light curves. In Sect.~\ref{sec:dis} we discuss the implications of
our results in the broader context of young eruptive stars.

%-----------------------------------------------------------------
% OBSERVATIONS
%-----------------------------------------------------------------
\section{Observations and data reduction}
\label{sec:obs}

\subsection{Radial velocity measurements}
\label{sec:spec}

\paragraph{FEROS.} We obtained 57 spectra of EX~Lup between 2007 July
and 2012 July with FEROS, the Fiber-fed Extended Range Optical
Spectrograph \citep{kaufer1999}, which is an \'echelle spectrograph
mounted at the MPG/ESO 2.2\,m telescope at La Silla Observatory,
Chile. The spectra cover the 3500--9200\,\AA{} wavelength range with
high resolution (R = 48\,000), distributed in 39 different \'echelle
orders. The wavelength calibration was done with a thorium-agron
(ThAr) lamp and spectra were obtained in object-cal (simultaneous
exposure to the ThAr lamp during the target observation) and in
object-sky mode (simultaneous sky exposure). Exposure times were
typically 1200 -- 1500\,s, but in some cases were increased to 3000\,s
because of bad weather conditions. The spectra were reduced using the
online Data Reduction System (DRS) on site, which included the
following steps: detection of spectral orders, wavelength calibration,
background subtraction, flatfield correction, and order
extraction. Finally, the DRS rebinned the reduced spectra to constant
wavelength steps and merged the individual orders. The signal-to-noise
ratio in the final spectra is between 5 and 80. Given the known
problem with the barycentric correction provided by the FEROS
pipeline \citep{muller2013}, we recalculated the barycentric
correction by using the IDL routine baryvel.pro\footnote{The
    barycentric correction in the FEROS pipeline was found to be
    inaccurate, which introduces an artificial one-year period with a
    semi-amplitude of 62\,m\,s$^{-1}$. The algorithm used in
    baryvel.pro is accurate to $\sim$1\,m\,s$^{-1}$. For details, see
    \citet{muller2013}}. The necessary corrections were in the range
of 0--70 m/s, with a seasonal variation.

\paragraph{HARPS.} We obtained ten spectra of EX~Lup between 2008 May
and 2009 March with HARPS, the High Accuracy Radial velocity Planet
Searcher \citep{mayor2003}, which is a fiber-fed high-resolution (R =
115\,000) spectrograph mounted at the 3.6\,m telescope at La Silla
Observatory, Chile. The instrument is able to obtain target and ThAr
calibration spectra simultaneously, as well as target and sky spectra
at once. The spectra cover the 3780--6910\,\AA{} wavelength range,
distributed over the \'echelle orders 89--161. Exposure times were
between 600\,s and 1200\,s. The reduced data were obtained from the
ESO archive pipeline processed data
query\footnote{archive.eso.org/wdb/wdb/eso/repro/form}. The spectra
have similar signal-to-noise ratios to the FEROS spectra.

\paragraph{RV determination.} Because EX~Lup is a highly
active star, special care had to be taken when determining its
RV. During the 2008 outburst, except for a very veiled Li 6708\AA{}
line, no photospheric absorption lines were visible in the optical and
near-infrared spectra of EX~Lup \citep{sicilia2012, kospal2011a}.
Thus RV determination was not possible from those spectra. Before the
outburst in 2007, and after the outburst in 2009--2012, however,
several photospheric absorption lines were present. We used the list
of emission lines detected by \citet[][see their Table
  2]{sicilia2012}, and checked whether we see emission at these
wavelengths, either as pure emission lines, or as little narrow
emission peaks superimposed on broader absorption lines. The affected
absorption lines were discarded from the RV analysis.

To obtain RV from the absorption lines (i.e., the RV of the central
star), we cross-correlated the object spectra with a synthetic
template spectrum. We chose a synthetic template with an effective
temperature of $T_{\rm eff} = 3750\,$K, surface gravity of $\log g =
4.0$, and solar metallicity, because EX~Lup is an M0.5-type star with
an estimated stellar mass of $M_*=0.6$\,M$_{\odot}$ and stellar radius
of $R_*=1.6$\,R$_{\odot}$
\citep{grasvelazquez2005,sipos2009,aspin2010}. We synthesized the
template spectrum using the SPECTRUM
software\footnote{http://www1.appstate.edu/dept/physics/spectrum/spectrum.html}
by \citet{gray1994}. Following the method of \citet{reiners2012}, we
broadened the synthetic spectrum to account for the line broadening
due to the stellar rotation in the object spectrum. Owing to the
finite spectral resolution of FEROS (and the resulting instrumental
broadening of about 2-3\,km\,s$^{-1}$), we could only determine an
upper limit of $v \sin i < 3$\,km\,s$^{-1}$ for EX~Lup. We obtained
the same upper limit for $v \sin i$ from the HARPS spectra, which is
consistent with the detection limit of HARPS without observing a large
sample of M dwarfs (cf.~\citealt{houdebine2010,reiners2012}).

We used similar RV determination procedures for both FEROS and
HARPS. In the case of FEROS, we used 10 \'echelle orders between
5580\,\AA{} and 7875\,\AA{}, and discarded shorter wavelengths because
the signal-to-noise ratio was too low. We avoided regions contaminated
by telluric absorption features. We calculated the cross-correlation
function (CCF) of the observed and the synthetic spectrum for each
individual order separately. By fitting a Gaussian to the CCF, we
determined the radial velocity shift for each order. We checked
whether there is any systematic difference in the RV values obtained
for bluer or redder orders, but detected no difference. Thus, we
calculated a weighted average of the RVs from all orders. The errors
for the average values were calculated using the relation derived in
\citet{setiawan2003}. In the case of HARPS, we used nine different
orders between 5450\,\AA{} and 6865\,\AA{}, again avoiding telluric
absorption lines, and determined the RV and its uncertainty the same
way as for FEROS, using the same CCF template. In total we obtained 45
FEROS and 9 HARPS RV values. Five FEROS spectra were unusable because
of very low signal-to-noise ratio caused by bad weather or high
airmass, while seven FEROS spectra and one HARPS spectrum had to be
discarded because they were taken in 2008 when EX~Lup was in
outburst. The results are listed in Table~\ref{tab:rv}.

The quiescence spectra of EX~Lup are very rich in emission
lines. Such lines are assumed to form in the hot gas within the
accretion columns \citep[e.g.,][]{beristain1998}, and given that they
appear at near-zero velocity, they probably originate from hot gas not
far from the stellar photosphere. To understand the causes of the
EX~Lup RV variations, we also measured separate RVs for the emission
lines. For this purpose, we selected a number of strong, narrow (FWHM
typically around 10-20\,km\,s$^{-1}$) emission lines in the quiescent
spectra of EX~Lup. We tested three different sets of lines: (a) 60
emission lines identified in the one quiescent spectrum with the best
signal-to-noise ratio; (b) 133 emission lines identified by
\citet{sicilia2012} in the outburst spectra; and (c) 25 Fe\,II lines,
the strongest ones among the emission lines visible in the quiescent
spectra (a subset of the 133 lines in point b). We used the
RVSAO/EMSAO package in IRAF to determine RV from the emission lines by
fitting Gaussians to the detected narrow emission lines. We found that
the RVs obtained with the different sets of lines were consistent
within the measurement uncertainties, although there is a hint of
small systematic differences between the ionized and the neutral
lines. In the following, we will use the emission line RVs detemined
in point (b) because these have the smallest uncertainties. These
results are also listed in Table~\ref{tab:rv}.

\subsection{Optical and infrared light curves}

\paragraph{Spitzer.} We obtained 3.6$\,\mu$m and 4.5$\,\mu$m
observations of EX~Lup with IRAC, the infrared camera on-board the
Spitzer Space Telescope, as part of a post-He program aimed at
monitoring low- and intermediate mass young stellar objects (PID:
60167, PI: P. \'Abrah\'am). Observations were taken between 2010 April
24 and May 7 with an approximately one-day cadence. On each day,
images were taken in full array mode, using an exposure time of 0.2 s
and a 5-point dithering. We downloaded corrected basic calibrated data
(CBCD) produced by the pipeline version S18.18 from the Spitzer
Archive. Photometry was made on individual CBCD frames that were
corrected for array location dependence. We performed aperture
photometry with an aperture radius of 2\farcs4 (2 pixels), and sky
annulus between 14\farcs4 and 24{\arcsec} (12 and 20 pixels). For the
sky estimates, we used an iterative sigma-clipping method with a
clipping threshold of 3$\sigma$. We applied pixel phase correction to
the 3.6$\,\mu$m data. Aperture corrections were 1.205 at 3.6$\,\mu$m
and 1.221 at 4.5 $\mu$m (IRAC Instrument Handbook v2.0). Finally, we
computed the average and rms of the individual flux values measured in
the different dither positions (at each band). The obtained fluxes and
their uncertainties can be found in Table~\ref{tab:photo}.

\paragraph{REM.} Ground-based optical and near-infrared observations
were obtained with the 60\,cm diameter Rapid Eye Mount (REM) telescope
located in La Silla, Chile \citep{covino2004}. The telescope is
equipped with two cameras that are operated simultaneously. Optical
images using the $V$ filter were taken with the ROSS instrument, while
infrared images using the $J$, $H$, and $K^{\prime}$ filters were
taken with the REMIR instrument: ROSS has an 1024$\times$1024 pixel
Apogee Alta CCD camera with a pixel scale of 0$\farcs$575 and REMIR
has a 512$\times512$ Rockwell Hawaii detector with a pixel scale of
1$\farcs$2. Both cameras have a field of view of about
10$'{\times}$10$'$. Observations were executed in service mode during
14 nights between 2010 April 24 and May 9, in most cases
simultaneously with the Spitzer observations.

Nine $V$-band images were obtained on each night with an exposure time
of 60\,s. We corrected the images for bias, dark, and flatfield in the
usual way. Then, we shifted and co-added the nine frames in order to
increase the signal-to-noise ratio. We obtained aperture photometry on
the co-added frames using an aperture radius of 4$\farcs$6 (8 pixels)
and a sky annulus between 17$\farcs$25 and 23$''$ (30 and 40
pixels). We calculated instrumental magnitudes for the science target
and four other stars in the field, to be used as comparison stars in
differential photometry. For absolute calibration of the magnitude
scale, V magnitudes of four comparison stars (NOMAD 0496-0417325,
0496-0417355, 0496-0417185, and 0496-0417090) were taken from the
NOMAD catalog \citep{nomad}.

The near-infrared images were obtained with exposure time of 5\,s (in
the $H$ and $K^{\prime}$ filters) or 10\,s (in the $J$ filter). Each
observation was taken at five dither positions, which were combined to
eliminate the sky signal and correct for flatfield differences. We
performed aperture photometry on the individual dither frames, with a
radius of 3$\farcs$6 (3 pixels) and a sky annulus between 12$''$ and
16$''$ (10 and 13 pixels). For the photometric calibration, we used
the Two Micron All Sky Survey (2MASS) catalog \citep{cutri2003}. To
avoid any remaining flatfield inhomogeneities within the image, we
used the two closest bright 2MASS stars with AAA quality flag
(2MASS16031144-4018178, and 2MASS 16030056-4018290) as comparison
stars. We determined the offset between the instrumental and the 2MASS
magnitudes by combining the five dither measurements of both
comparison stars using an outlier-resistant algorithm. No color term
was needed in this transformation. The uncertainty of the final
photometry is the quadratic sum of the formal uncertainty of the
aperture photometry, the scatter of the photometry of the individual
dither frames, and photometric calibraion. The resulting photometry is
listed in Table~\ref{tab:photo}.

%-----------------------------------------------------------------
% RESULTS AND ANALYSIS
%-----------------------------------------------------------------
\section{Results and analysis}
\label{sec:res}

\begin{figure}
\centering
\includegraphics[scale=0.47]{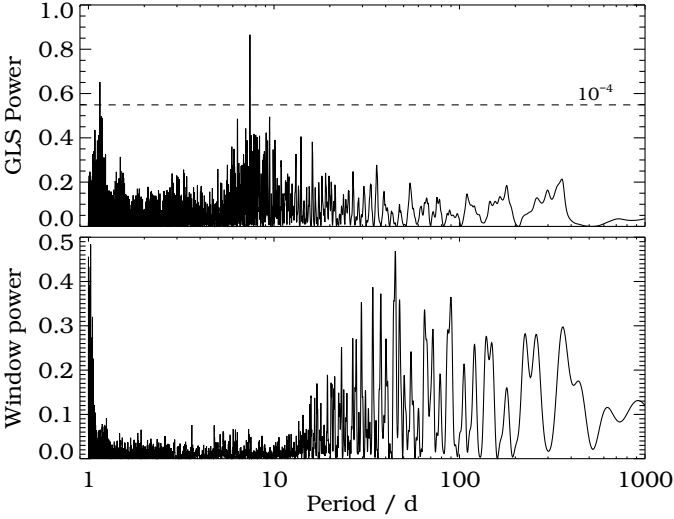}
\caption{{\it Top:} GLS periodogram of the FEROS RV values, showing
  the best period of 7.417\,days. The horizontal dashed line indicates
  the FAP level of $10^{-4}$. {\it Bottom:} window function.}
\label{fig:GLS}
\end{figure}

\begin{figure}
\centering
\includegraphics[scale=0.58]{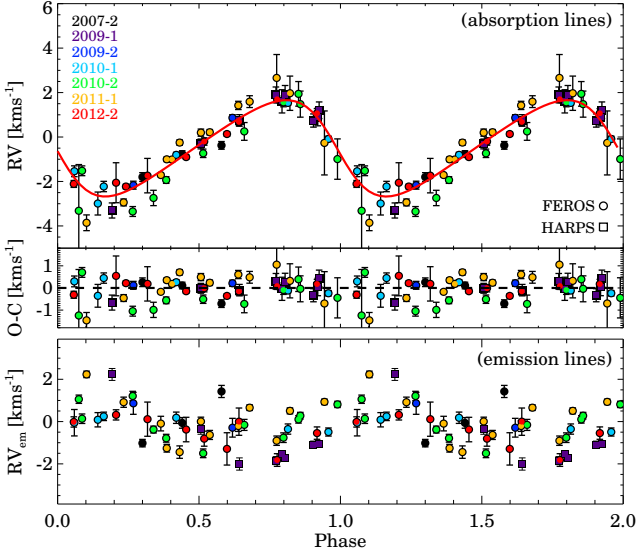}
\caption{{\it Top:} Best Keplerian fit to the absorption line RVs for
  the combined FEROS and HARPS RV data set (with RVlin). The different
  colors indicate different observing seasons (--1 stands for the
  first half of the year, --2 for the second half). Circles represent
  FEROS and squares HARPS data points. {\it Middle:} residua with the
  same color and shape coding. {\it Bottom:} Emission line RVs with
  the same color and shape coding.}
\label{fig:folded}
\end{figure}

\subsection{Period analysis of the RV data and orbital solution}
\label{sec:kepler}

To search for periodicities in the RV values measured from the
absorption lines, we used the generalized Lomb-Scargle (GLS) method
\citep{zechmeister2009}. Since the number of available HARPS points is
not sufficient to carry out an independent RV period determination, we
first considered the FEROS data only, and afterwards we combined the
FEROS and HARPS data points, and repeated our analysis. The data
points were weighted with the inverse square of their
uncertainties. Figure~\ref{fig:GLS} shows the GLS periodogram
calculated for the 45 FEROS points. Two significant periods with false
alarm probability (FAP) of $<$\,$10^{-4}$ are visible, one at around
1\,day, which is an alias caused by the sampling (see the window
function in the bottom panel of Fig.~\ref{fig:GLS}), and another one
at 7.417\,days. The RV curve phase-folded with the 7.417-day period is
plotted in Fig.~\ref{fig:folded}.

Figure~\ref{fig:folded} exhibits a clear periodic signal with an RV
semi-amplitude of $\approx$2.2\,km\,s$^{-1}$. The shape of the curve
is asymmetric, with the increasing part approximately 1.7 times longer
than the decreasing part. In the figure, we color-coded the different
observing seasons during our five-year monitoring program. No
systematic differences can be seen between the different seasons,
which suggests that the period, phase, and amplitude of the RV
variations were stable for at least four years, between 2009 and
2012. The three data points from 2007 are also consistent with the
later observations, implying that the RV variations were largely
unaffected by the largest outburst of EX~Lup ever observed which
occurred in 2008.

The shape of the phase-folded RV curve suggests the possibile
existence of a companion on an eccentric orbit around EX~Lup. We
fitted a Keplerian solution to the FEROS RVs using both GLS and the
idl code RVlin\footnote{http://exoplanets.org/code/}
\citep{wright2009}. The resulting parameters are listed in
Table~\ref{tab:keplerfitparameters}. The RVlin code can handle data
from different instruments by fitting parameters to correct for the
relative offset between them, thus, we used it to fit the combined
FEROS+HARPS data sets, and determined an offset of 0.291\,km\,s$^{-1}$
between the two instruments. The code also calculates uncertainties
for the fitted parameters using a bootstrapping algorithm
\citep{wang2012}. The orbital parameters determined from the fit for
the combined data sets are also listed in
Table~\ref{tab:keplerfitparameters}. The values derived for the
different data sets using different methods are all consistent within
the uncertainties.

To calculate $m \sin i$ and the semi-major axis $a$, we assumed a
stellar mass of $M_*=0.6$\,M$_{\odot}$ for EX~Lup
\citep{grasvelazquez2005}. Figure~\ref{fig:folded} shows the best fit
for the combined data set, which has a FAP of
6.7$\times$10$^{-27}$. After subtracting the 7.417-day period
Keplerian fit from the RV data, we searched for sine periodicities in
the radial velocity residuals, but found no significant periods with
FAP $<$\,$10^{-4}$.

As described in Sect.~\ref{sec:spec}, we calculated RVs from those
narrow emission lines that are isolated and not superimposed on
broader absorption features. Similar but weaker emission lines may
distort the absorption lines and might introduce artificial RV
signals, despite our best efforts to discard the affected absorption
lines from our analysis. To make sure that this is not the case, we
performed a period analysis on the RVs of the emission lines. The
algorithm found no significant periods, although a small peak is
present at about 7\,days with FAP of 10$^{-2}$. In
Fig.~\ref{fig:folded}, we plotted the RVs of the emission lines phase
folded with the 7.417\,d period. The data points have a large scatter
compared to the RVs determined from the absorption lines, and they
show insteead a sinusoidal variation with a semi-amplitude of about
0.8\,km\,s$^{-1}$. From this exercise, it is clear that emission
components moving at this velocity cannot cause the RV signal observed
in the absorption lines.

\begin{table}
\centering
\begin{tabular}{lcc}
\hline\hline
Parameter                   & Unit		 & Value		   \\
\hline
Period                      & day		 & 7.417 $\pm$ 0.001	   \\
RV semi-amplitude           & km\,s$^{-1}$	 & 2.18 $\pm$ 0.10	   \\
Eccentricity                & $\dots$		 & 0.23 $\pm$ 0.05	   \\
Longitude of periastron     & $^{\circ}$	 & 96.8 $\pm$ 11.4	   \\
Epoch of periastron passage & JD		 & 2455405.112 $\pm$ 0.224 \\
RV offset                   & km\,s$^{-1}$	 & --0.52 $\pm$ 0.07	   \\
FAP $^1$                    & $\dots$		 & 6.7$\times$10$^{-27}$   \\
\hline
$m \sin i$ $^2$             & $M_{\rm{Jupiter}}$ & 14.7 $\pm$ 0.7	   \\
semi-major axis $^2$        & AU		 & 0.063 $\pm$ 0.005	   \\
\hline
\end{tabular}
\caption{Best-fitting parameters of the Keplerian solution fitted to
  the RV measurements of EX~Lup (with RVlin, using the combined FEROS
  and HARPS dataset). ($^1$): false alarm probabilty of the Keplerian
  fit as defined by \citet{cumming2004}. ($^2$): assuming a stellar
  mass of 0.6\,M$_{\odot}$.}
\label{tab:keplerfitparameters}
\end{table}

\subsection{Analysis of the stellar activity}
\label{sec:activity}

The presence of a companion is not the only possible explanation for
periodic RV changes. Photospheric stellar activity, i.e., dark (cold)
or bright (hot) spots, are known to result in periodic RV variations
\citep[e.g.,][and references therein]{lanza2011}. For instance, a cool
spot on the stellar surface would produce a deficit of emission at the
spot's velocity within the line profile, and the rotation of the star
then leads to an RV signal which is periodic with the rotational
period. Additionally, RV measurements from lines with different
temperature sensitivities would yield different RV amplitudes
\citep[e.g.,][]{hatzes1999}. Thus, by analyzing the distortion of the
spectral lines and their correlation with the RV, or by analyzing line
ratios sensitive to the effective temperature, it is possible to
verify the presence of stellar spots. In addition to the photospheric
spots, cool stars, especially the rapidly rotating younger ones, have
very active chromospheres \citep{montes2004}. The chromospheric
spectrum is dominated by emission lines, which again can distort the
photospheric absorption line profiles, or in extreme cases can even
fill them up or turn them into emission lines. Chromospheric activity
may increase the noise of RV measurements, but it may also produce
periodic RV signals over a few rotational periods
\citep{santos2003}. In the following, we check whether the RV
variations and their periodicity observed for EX~Lup can be due to
stellar activity.

\begin{figure}
\centering
\includegraphics[angle=0,scale=0.62]{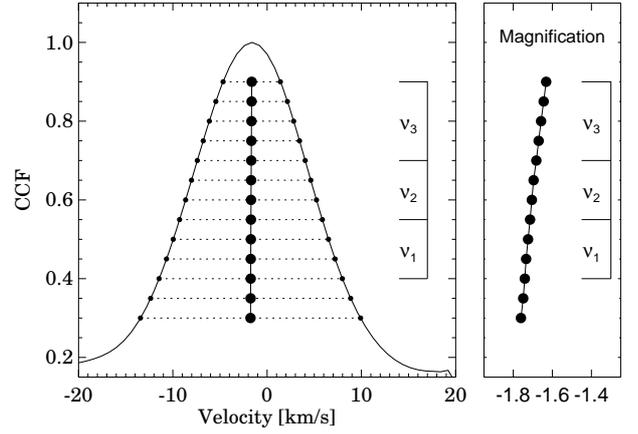}
\caption{Bisector of the CCF computed from the FEROS spectrum of
  EX~Lup observed on 2010 July 23.}
\label{fig:bisector}
\end{figure}

\paragraph{Bisector analysis.}
The CCF represents the mean absorption line profile of the star. We
computed the CCF for each spectrum by combining the individual CCFs of
the different orders into a master CCF using a robust averaging method
that assigns less weight to the deviating orders. One way to analyze
its distortion is to compute the bisector
\citep[e.g.,][]{queloz2001,gray2005}, as illustrated in
Fig.~\ref{fig:bisector}. Following \citet{dall2006} we computed the
mean bisector velocities for different regions in the CCF, and
estimated the formal uncertainty of the mean values from the
dispersion of the bisector points within that region.  Following
Povich et al. (2001), we defined three regions: $v_1$ for 0.4 $\le$
CCF $\le$ 0.55, $v_2$ for 0.55 $\le$ CCF $\le$ 0.7, and $v_3$ for 0.7
$\le$ CCF $\le$ 0.9 (Fig.~\ref{fig:bisector}). These CCF ranges were
suitable for most spectra; however, in a few cases the continuum level
was high and affected the lowest bisector points in the $v_1$
range. These measurements were discarded from further analysis. In the
next step, we calculated the bisector velocity span (BVS\,$=v_3-v_1$),
the bisector curvature (BC\,$=(v_3-v_2) - (v_2-v_1)$), and the
bisector velocity displacement (BVD\,$=(v_1+v_2+v_3)/3-\lambda_c$,
where $\lambda_c$ is the observed central wavelength).

\begin{figure}
\centering
\includegraphics[angle=0,scale=0.67]{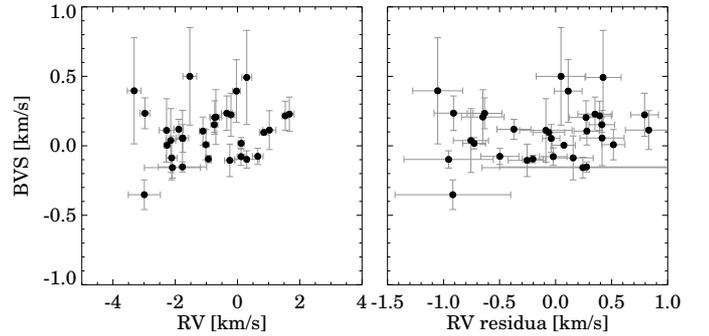}
\caption{BVS values plotted against the RV ({\it left}) and against
  the RV residua ({\it right}).}
\label{fig:bisector_corr}
\end{figure}

\begin{figure}
\centering
\includegraphics[angle=-90,scale=0.37, trim = -10mm 0mm -15mm 0mm,
  clip]{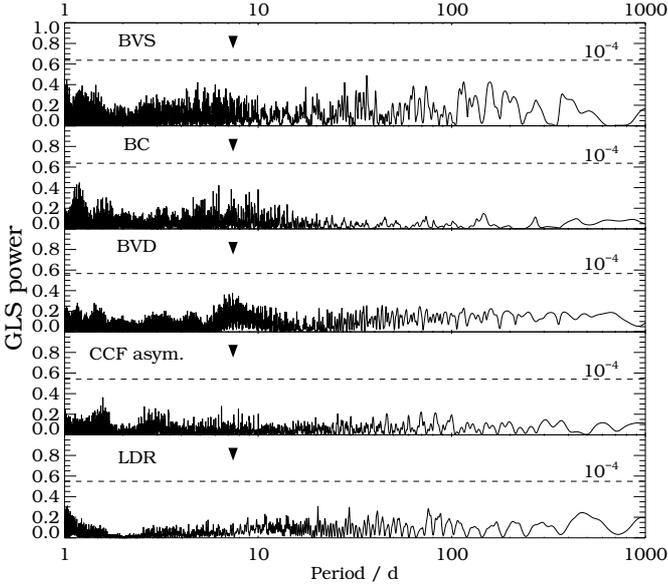}
\caption{GLS periodograms of the BVS, BC, BVD, CCF asymmetry, and the
  LDR of V\,I and F\,I. The arrows mark the 7.417-day RV period. The
  horizontal dashed lines indicate a FAP level of $10^{-4}$.}
\label{fig:bisector_gls}
\end{figure}

We checked whether the BVSs correlate with the RV, which is expected
if the RV variations are caused by rotational modulation due to
starspots \citep[e.g.,][]{queloz2001}. Figure~\ref{fig:bisector_corr}
shows the BVS plotted as a function of the RV and the RV residua
(after the subtraction of the Keplerian solution). The linear Pearson
correlation coefficients \citep{pearson1920} for the two graphs are
0.08 and 0.27, respectively, indicating no correlation between the
quantities. As an independent check, we also performed a bisector
analysis on carefully selected individual absorption lines in the
FEROS spectra. Similarly to our previous results, no correlation
between the BVS and the RV was found. We calculated GLS periodograms
for the BVS, BC, and BVD (Fig.~\ref{fig:bisector_gls}), but found no
significant period with FAP smaller than 10$^{-4}$. As indicated by
the arrows in the figure, no peak is present at a period of 7.417\,day
for the BVS, BC, and BVD.

\paragraph{Asymmetry of the CCF.} Distortions of the CCF profile can
also be analyzed by first plotting the slope of the CCF at each
velocity, then by integrating over the velocity axis with positive and
negative slopes, respectively, and calculating the ratio of the
positive and negative areas. For a perfectly symmetric Gaussian, this
ratio would be 1, while for a distorted Gaussian, it would be below or
over 1. We determined this measure of the CCF asymmetry for each order
separately, then took their weighted average. Finally, we calculated
the GLS periodogram (Fig.~\ref{fig:bisector_gls}), but again found no
significant periods, and no peak at a period of 7.417\,days.

\begin{figure}
\centering
\includegraphics[angle=-90,scale=0.36]{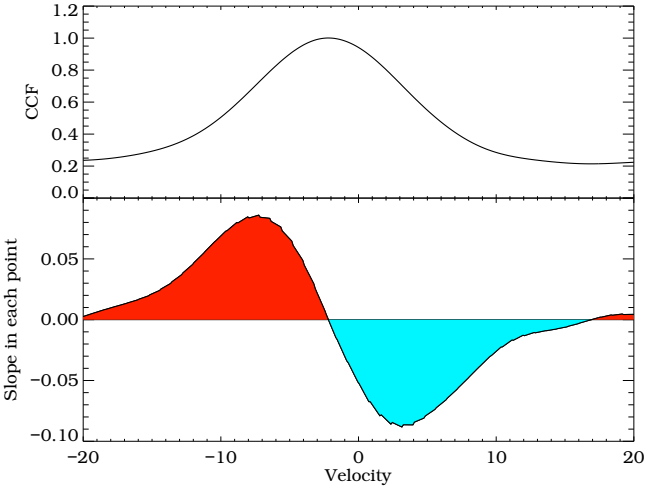}
\caption{Illustration of the calculation method of the CCF
  asymmetry. The quantity of asymmetry is determined by obtaining the
  quotient of the area of positive slope (red) and of negative slope
  (blue). The CCF displayed here was computed from the FEROS spectrum
  of EX~Lup observed on 2010 July 23.}
\label{fig:CCF_asym_EX_Lup}
\end{figure}

\paragraph{Temperature-sensitive spectral features.} A cool spot on
the stellar surface, rotating in and out of view, changes the
effective temperature of the visible stellar hemisphere. This has an
influence on the strength of temperature-sensitive absorption lines or
bands. For instance, \citet{catalano2002} showed that both the
6268.87\,\AA{} V\,I line and the 6270.23\,\AA{} Fe\,I line become
stronger with decreasing temperature, but the variation of the line
depth of the low-excitation V\,I line is more pronounced than that of
Fe\,I. For this reason, the line depth ratio (LDR) is a good tracer of
changes in the effective surface temperature. We calculated the GLS
periodogram of the LDR of V\,I and Fe\,I
(Fig.~\ref{fig:bisector_gls}), but found no significant period, and no
peak at 7.417\,days.

In the presence of a cool spot, bands like TiO, CaH, and CaOH may
appear or strengthen \citep[see][who measured these bands in a large
  number of M dwarfs]{reid1995}. Using our FEROS spectra, we
determined the TiO 1, TiO 2, TiO 3, TiO 4, TiO 5, CaH 2, CaH 3, CaOH,
and H$_{\alpha}$ indices as defined by \citet{reid1995}, and looked
for periodic changes due to spots on the stellar surface. Similarly to
the other stellar spot indicators discussed above, we found no
significant period in the GLS periodograms of these spectral indices,
and no peak at a period of 7.417\,days.

\paragraph{Analysis of the Ca lines.} \citet{larson1993} have shown that
calcium lines in the stellar spectrum, such as the Ca\,II H and K
lines at 3968\,\AA{} and 3933\,\AA{} and the Ca\,II infrared triplet
at 8498\,\AA{}, 8542\,\AA{}, and 8662\,\AA{} are good indicators of
stellar chromospheric activity. Since the Ca\,II H line is possibly
blended with H$_{\epsilon}$, and the Ca\,II 8498\,\AA{} and
8542\,\AA{} lines are contaminated by terrestrial water vapor lines,
we discarded them from further analysis and concentrate on the Ca\,II
K line at 3933\,\AA{} and on the Ca\,II line at 8662\,\AA{}. For
Ca$\lambda$3933\,\AA{} we calculated the chromospheric activity index
$S_{\rm{FEROS}}$ and monitored its variation over
time. $S_{\rm{FEROS}}$ is defined as follows:
\begin{equation}
S_{\rm{FEROS}} = \frac{F_e}{F_b + F_r} = \frac{F_{\rm{3933}\,\AA~-
    \rm{3935}\,\AA}}{F_{\rm{3930}\,\AA~- \rm{3933}\,\AA} +
  F_{\rm{3935}\,\AA~- \rm{3938}\,\AA}},
\end{equation}
where $F_e$, $F_b$, and $F_R$ are the fluxes integrated over the
wavelength ranges indicated in the equation above. These S-indices for
the Ca\,II H and K lines were first defined by \citet{vaughan1978},
and they measure the strength of the line emission relative to the
adjacent continuum, an established chromospheric activity indicator
\citep[e.g.,][]{mittag2013}. Additionally we calculated the equivalent
width (EW) of Ca$\lambda$8662\,\AA{}, and applied the method of the
asymmetry analysis of the CCF to the emission core of the Ca\,II K
line. We calculated GLS periodograms for all three indicators, but
found no significant peak at the RV period of 7.417\,d. The
$S_{\rm{FEROS}}$ index and the Ca$\lambda$8662\,\AA{} line EW show a
peak around 1.0\,d with a FAP smaller than $10^{-5}$, which we
interpret as a feature of the sampling effect. The
Ca$\lambda$8662\,\AA{} line EW reveals a strong peak at a FAP smaller
than $10^{-5}$ at about 25\,d. However, by looking at the window
function of the EW values, it is apparent that this is a result of the
sampling as well. The Ca\,II K asymmetry does not exhibit any
significant peaks between 1 and 1000\,d. We note that the Ca\,II
infrared triplet is often considered a good accretion tracer
\citep[e.g.,][]{muzerolle1998}. EX~Lup has a non-negligible accretion
rate even in quiescence, and the broad wings of the Ca\,II lines
(Fig.~3 in \citealt{sicilia2012}) suggest that part of the line flux
is related to accretion rather than to chromospheric activity.

\subsection{Spot model}
\label{sec:spotmodel}

In all our previous analyses we found no evidence of periodic stellar
activity. Nevertheless, in the following we will assume that the
7.417\,d period we found in the RVs is in fact due to stellar
rotation, and we will try to find a spot model that can reproduce the
observed $\approx$2.2\,km\,s$^{-1}$ semi-amplitude of the RV
curve. Since there is no flat section in the RV curve, the spot should
practically always be visible to a certain extent. Given the large RV
semi-amplitude compared to the small $v \sin i < 3$\,km\,s$^{-1}$, we
expect spots with large filling factors. We constructed a simple spot
model and simulated a grid of single spots with different
temperatures, sizes, and latitudes on the photosphere of EX~Lup with
different stellar inclinations. The results showed that typical cold
or hot spots, covering a few percent of the stellar surface and having
a temperature difference of a few hundred K, are unable to reproduce
the measured RV semi-amplitude. Thus, we had to explore more extreme
parameters in our modeling. We found that a large cold spot covering
practically a whole hemisphere, with a filling factor between 80\% and
100\%, with a temperature of 1500\,K -- 2500\,K cooler than the
stellar photosphere at low latitudes (within 30$^{\circ}$ of the
equator), and with low stellar inclinations (the angle between the
line of sight and the stellar equator being between 0 and
30$^{\circ}$) reproduced the observed RV semi-amplitude. Similarly,
even if it is unrealistic, the only hot spot that produced an adequate
fit to the RV semi-amplitude covered a whole hemisphere with a
temperature of 10\,500\,K hotter than the photosphere at a latitude of
0$^{\circ}$, and an inclination of 0$^{\circ}$.

\begin{figure}
\centering \includegraphics[width=\columnwidth,angle=0]{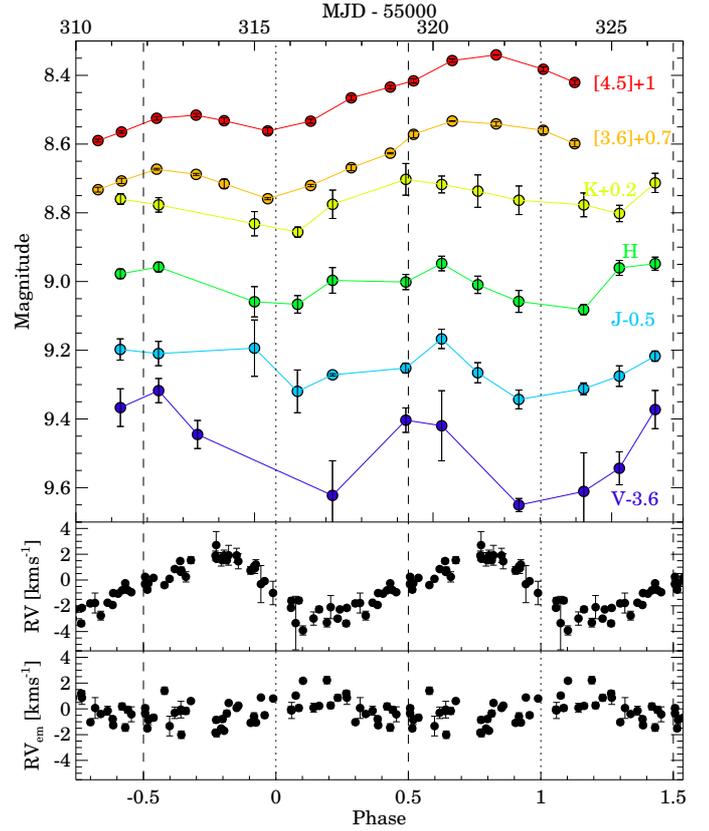}
\caption{Light curves and RV curves of EX~Lup. For clarity, the light
  curves were shifted along the y-axis by the amounts indicated on the
  right side. Dashed and dotted lines mark when the RV equals the
  systemic velocity.}
\label{fig:light}
\end{figure}

To test if any of these spot models could reproduce the observed
photometric behavior, we plotted in Fig.~\ref{fig:light} the light
curves of EX~Lup between 0.55 and 4.5$\,\mu$m covering about two
weeks in 2010 with an approximately daily cadence. We observed
significant brightness changes in this period at all wavelengths. The
curves between $V$ and $H$ exhibit similar shapes with a decreasing
amplitude towards longer wavelengths. The peak-to-peak amplitude of
the variability is $\Delta{}V=0.33\,$mag, $\Delta{}J=0.18\,$mag, and
$\Delta{}H=0.14\,$mag. The curves seem to be periodic with minima
around MJD = 55316 and MJD = 55323, which is consistent with the
rotation period of our spot model. This periodicity is observable also
between $K$ and [4.5], but superimposed on a rising trend. There is a
hint of a phase lag in the maxima towards longer wavelengths. The
peak-to-peak variability amplitudes here are $\Delta{}K=0.16\,$mag,
$\Delta{}[3.6]=0.24\,$mag, and $\Delta{}[4.5]=0.25\,$mag, as measured
directly on the light curves, and $\Delta{}K=0.14\,$mag,
$\Delta{}[3.6]=0.17\,$mag, and $\Delta{}[4.5]=0.13\,$mag if we
subtract a linearly increasing trend.

\begin{figure}
\centering
\includegraphics[angle=0,scale=0.6]{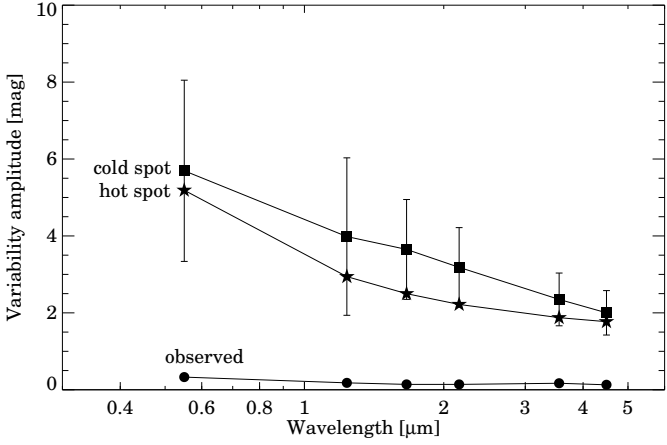}
\caption{Variability amplitudes as measured on the light curves of EX
  Lup and as obtained from the spot models made to reproduce the
  observed RV semi-amplitude (for details, see
  Sect.~\ref{sec:spotmodel}).}
\label{fig:ampl}
\end{figure}

We calculated the photometric variability amplitudes from our spot
models using the limb darkening coefficients from John Southworth's
JKTLD code\footnote{http://www.astro.keele.ac.uk/jkt/codes/jktld.html}
(0.787 for the $V$ band, 0.470 for the $J$ band, 0.439 for the $H$
band, 0.366 for the $K$ band, 0.190 at 3.550$\,\mu$m, and 0.142 at
4.493$\,\mu$m). The obtained values (the median of several models that
were all consistent with the observed RV amplitude, with the standard
deviation as error bars) are plotted in Fig.~\ref{fig:ampl}. It is
evident that the amplitudes for the spot models exceed the observed
ones by several magnitudes. In the case of a cold spot, the phase when
the spot faces towards the observer corresponds to the lowest
photometric brightness and the point in the RV curve when it becomes
smaller than the systemic velocity (dotted lines in
Fig.~\ref{fig:light}). Conversely, in the case of a hot spot, the
phase when the spot faces towards the observer corresponds to the
highest photometric brightness and the point in the RV curve when it
becomes larger than the systemic velocity (dashed lines in
Fig.~\ref{fig:light}). While this is approximately consistent with our
observations, the extremely large variability amplitudes predicted by
the spot models still make it improbable that spots are wholly
responsible for the observed RV variations.

In case spots cause the RV variations of EX~Lup, the observed RV
semi-amplitude should depend on the wavelength. As the contrast
between the spot and the unspotted stellar photosphere decreases with
increasing wavelength, we expect gradually smaller RV amplitudes. For
our large cold spot model, the difference between 5580\,\AA{} and
7875\,\AA{} (roughly corresponding to the bluest and reddest FEROS
orders we used) would be 0.05\,km\,s$^{-1}$ in the RV semi-amplitude.
For our large hot spot model, the value would be
0.28\,km\,s$^{-1}$. As we briefly mentioned in Sect.~\ref{sec:spec},
we found no significant difference in the RVs obtained from bluer or
redder orders. Our RV data allowed us to determine the RV
semi-amplitude to a precision of about 0.1\,km\,s$^{-1}$
(Table~\ref{tab:keplerfitparameters}). Thus, while we cannot exclude
the cold spot scenario based on these arguments alone, the hot spot
scenario seems to be unlikely.

Finally, we checked whether we can find a spot model (either cold or
hot) that would reproduce the observed photometric variability
amplitudes and calculated the expected RV semi-amplitudes. We found
that a cold spot with a temperature of 1500\,K cooler than the
photosphere and covering 11\% of a hemisphere would give $\Delta
V$=0.33\,mag (both the latitude of the spot and the inclination of the
star was taken to be 0$^{\circ}$). The same $\Delta V$ can also be the
result of a hot spot with a temperature of 525\,K hotter than the
photosphere (covering factor, latitude, and inclination are the same
as before). However, these relatively small spots would only cause an
approximately 0.3\,km\,s$^{-1}$ RV semi-amplitude, much smaller than
the observed 2.2\,km\,s$^{-1}$. Thus, we can conclude that the
observed photometric and RV variations cannot be reproduced at the
same time with cold or hot spots on the stellar surface.

%-----------------------------------------------------------------
% DISCUSSION
%-----------------------------------------------------------------
\section{Discussion}
\label{sec:dis}

\subsection{EX~Lup: a spotted or active star?}

In Sects.~\ref{sec:activity} and \ref{sec:spotmodel} we investigated
in detail whether the observed RV signal could be caused by cold or
hot spots on the stellar surface. We verified that spot indicators
like the BVS, BC, and CCF asymmetry do not exhibit any
periodicity. Nor do we see any dependence of the observed RV on
wavelength within the range covered by the FEROS spectra. We
constructed a simple spot model to reproduce the observed RV
semi-amplitude. Our modeling suggests that the spot would have to
cover a complete hemisphere, which is an extreme and atypical
solution. Moreover, the temperature of this extended spot deviates
from that of the photosphere by several thousand K. Thus, unless it
cools mainly through line emission \citep[c.f.][]{dodin2013}, it would
would cause periodic photometric changes of 4-8\,mag in the $V$ band,
which is not seen in our observations. A spot model that reproduces
the observed $V$ band variability amplitude fails to reproduce the
large RV semi-amplitude. We note that in the middle of our observing
campaign (2008) EX~Lup produced its historically largest outburst when
the accretion rate increased by a factor of 30. This event probably
did not leave unchanged the size and location of the spots; however
this rearrangement is not evident from the data.

\begin{figure}
\centering
\includegraphics[scale=0.6,angle=0]{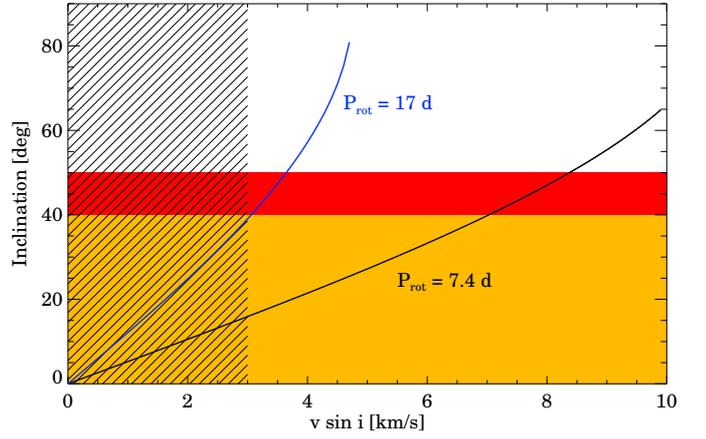}
\caption{Inclination vs.~$v \sin i$. The solid lines indicate the
  period of 7.4\,d (obtained from the RV analysis) and 17\,d (a lower
  limit obtained from the inclination constraint). The hatched area
  marks the $v \sin i$ constraint from the FEROS and HARPS spectra;
  the orange and red areas mark the constraints for disk inclination
  from \citet{sipos2009} and \citet{goto2011}, respectively.}
\label{fig:incl}
\end{figure}

In the stellar spot scenario, the 7.4\,d period of the RV curve would
represent the rotational period of the star. The true stellar
rotational period of EX~Lup is unknown. However, considering our
upper limit of 3\,km\,s$^{-1}$ for $v \sin i$, EX~Lup is probably not
a very fast rotator, except if the inclination of the star's equator
is sufficiently low. A rotation period of 7.4\,d would limit the
inclination below 16$^{\circ}$ (Fig.~\ref{fig:incl}). If the planes of
the circumstellar disk and the star's equator are not very different,
modeling the circumstellar disk geometry also provides information on
the inclination. \citet{sipos2009} fitted the broad-band spectral
energy distribution of the quiescent EX~Lup system, and were able to
constrain the disk inclination to be between 0$^{\circ}$ and
40$^{\circ}$. The lower part of this range is thus consistent with the
7.4\,d rotation (Fig.~\ref{fig:incl}). \citet{goto2011}, however,
found that a disk inclination between 40$^{\circ}$ and 50$^{\circ}$ is
needed to fit the fundamental vibrational lines of CO observed in the
4.5--5$\,\mu$m range during the 2008 outburst. The combination of the
two constraints give a most probable inclination of 40$^{\circ}$,
which, combined with our upper limit on $v \sin i$, indicates a
rotation period of at least 17\,d, inconsistent with the spot
scenario. We conclude that the explanation of the measured RV curve
with photospheric spots is an unlikely hypothesis.

We also checked in Sect.~\ref{sec:activity} whether flows and
inhomogeneities on the stellar surface or in the chromosphere could
produce the measured RV variations. In this scenario the RV period
again would be identical to the stellar rotation period. EX~Lup, as an
M-type star, is probably chromospherically active. However, the
phenomena responsible for the distortions of the line profiles and
thus for the RV variations are not expected to survive more than a few
rotation periods, while our measured RV curve, its phase, period, and
amplitude, was stable for at least five years. Moreover, our frequency
analysis revealed no periodic signal in the behavior of activity
indicators derived from the Ca\,II K line and from the Ca\,II infrared
triplet. Thus, although chromospheric activity might be present in EX
Lup and might add some random noise to the RV measurements, the
periodic signal is very unlikely to be due to the activity.

\subsection{Accretion scenarios in the quiescent EX~Lup system}

As we briefly discussed in Sect.~\ref{sec:activity}, EX~Lup displays
a number of narrow emission lines whose RV, phase-folded with a period
of 7.417\,d (Fig.~\ref{fig:light}, bottom), reveal slight sinusoidal
variations. Although the peak-to-peak amplitude is small and the
scatter is large, there seems to be an anti-correlation (or
180$^{\circ}$ phase shift) between the RV of the emission lines and
the RV of the absorption lines. EX~Lup is not the first object where
an anti-correlation of this kind has been observed: \citet{gahm1999}
discovered that RW\,Aur\,A shows photospheric RV variations with an
amplitude of 5.7\,km\,s$^{-1}$ and period of 2.77\,d, while the RV of
the narrow emission components of the He\,I and He\,II lines vary in
anti-phase with the photospheric lines. \citet{petrov2001} proposed
two possible interpretations. One possibility is that RW\,Aur\,A is a
single star whose rotational and magnetic axes are misaligned and the
magnetic poles are associated with the footprints of accretion
flows. The star in this model rotates with a period of 5.5\,d, and the
two active regions give rise to the observed 2.77\,d periodicity in
the RV of both the absorption and the emission lines \citep[see
  also][]{dodin2012}. The other scenario is that RW\,Aur\,A is a
binary with a brown dwarf companion on a 2.77\,d orbit. This companion
may induce a flow of material from one side of the inner disk to the
primary. The accretion flow could be generated either via
gravitational perturbations by the secondary, or by magnetospheric
interactions between the two components.

Our results in Sect.~\ref{sec:res} suggest a possible analogy between
RW\,Aur\,A and EX~Lup. Thus, in the following we will check whether
the explanations proposed by \citet{petrov2001} for RW\,Aur\,A can be
applied to EX~Lup as well. By analysing optical spectra taken during
the 2008 outburst, \citet{sicilia2012} concluded that there is a hot
and non-axisymmetric accretion flow or column(s) in the EX~Lup
system, where clumps of gas are accreted onto the star. Based on the
rapid recovery of the system after the outburst and the similarity
between the pre-outburst and post-outburst spectra, they also suggest
that the accretion channels are stable formations, and only the
accretion rate varies between the quiescent and outburst periods.

Following \citet{petrov2001}, it is possible that EX~Lup is a single
star with stable, rotating accretion column(s). A single column would
imply a 7.4\,d rotational period, while two opposite columns would
allow for a 14.8\,d period, which is more consistent with our
constraints on $v \sin i$ and inclination (Fig.~\ref{fig:incl}). The
structure of the accretion columns may resemble the one outlined by
\citet{dupree2012}, where the emission lines originate from an
accretion shock above the stellar surface. This picture could explain
the kinematics of the emission lines, and may cause photometric
variations as was observed in Fig.~\ref{fig:light}. The accretion
columns would distort the photospheric line profiles and cause an
apparent RV signal in the absorption lines. However, our observations
show no clear signature of line distortions (see
Figs.~\ref{fig:bisector}, \ref{fig:bisector_corr}, and
\ref{fig:CCF_asym_EX_Lup}).

In the alternative scenario, \citet{petrov2001} postulates a companion
that maintains an accretion column in the system. Our Keplerian fit to
the absorption line RVs in Sect.~\ref{sec:kepler} suggests that EX~Lup
has a close companion on an eccentric orbit with 7.4\,d period. In the
\citet{artymowicz1996} model, this companion may cause pulsed
accretion by periodically channeling material from the inner disk onto
the binary components with the same period as the binary orbital
period. In this picture, the accretion channel follows the orbital
motion of the companion, which would explain the RV curve of the
emission lines, while the absorption lines would wobble because of the
orbital motion of the binary. We note, however, that the situation may
actually be more complex. The typical magnetospheric truncation radius
of a T~Tauri star is about 7\,R$_*$ \citep{bouvier2007}. The binary
components orbit each other between 6.5\,R$_*$ and 10.4\,R$_*$, which
means that their magnetospheres partially overlap and may create a
complicated field for the accreting material (see Fig.~23 in
\citealt{petrov2001}).

In conclusion, both of the outlined pictures could work qualitatively
for EX~Lup, but the binary scenario accounts for more observed
properties. Further observations are necessary to prove the existence
of EX~Lup's companion. Nevertheless, in the following, we will
examine the properties of the hypothetical companion, and briefly
discuss its possible effect on the disk and the outbursts.

\subsection{The companion candidate of EX~Lup}

According to our Keplerian fit (Table~\ref{tab:keplerfitparameters}),
the $m \sin i$ of the companion candidate is approximately 15\,M$_{\rm
  Jup}$. Taking 50$^{\circ}$ as an upper limit for the inclination,
this gives a lower limit of 0.018\,M$_{\odot}$ for the mass of the
companion, putting it into the mass range of high-mass planets, brown
dwarfs, or very low-mass stars. According to the evolutionary models
of \citet{baraffe2002}, a brown dwarf with a mass of 0.02\,M$_{\odot}$
has an effective temperature of about 2500\,K (or higher if it is more
massive) for the first few million years of its life. This means a
spectral type of L0 or earlier and an absolute magnitude of
$M_J$=11\,mag or brighter (corresponding to J=17\,mag at a distance of
155\,pc), based on the results of \citet{dahn2002}. Since EX~Lup has
an average $J$=9.8\,mag, this is a brightness contrast of less than a
factor of 800.

The companion orbits EX~Lup between 0.049\,AU and 0.078\,AU (or
6.5\,$R_*$ and 10.4\,$R_*$). Interestingly, \citet{sipos2009} found
that the inner radius of the dust disk around EX~Lup in quiescence is
0.2\,AU, which is larger than the dust sublimation radius, and they
speculate that binarity may be responsible for clearing up the inner
circumstellar region. Simulations of binary-disk interactions indicate
that the inner radius of a circumbinary disk is typically 1.8--4 times
the semi-major axis of the binary for different eccentricities, mass
ratios, and disk viscosities \citep{artymowicz1994}. For the EX~Lup
system, this value is 2.6, well within the range predicted by the
simulations. Recent theoretical and observational results also support
the idea that companions can create an inner hole in the dust
distribution by trapping dust particles depending on grain size
\citep{rice2006,vandermarel2013}. Thus, if the existence of the
hypothetical companion proves to be true, it may have caused the
opening of the dust-free hole in the disk of EX~Lup.

Considering its small mass and separation, the companion candidate of
EX~Lup might be a unique object. The distribution of binary
separations peak at 2--16\,AU for main sequence M dwarfs
\citep{fischer1992,gizis2003}, at 30--50\,AU for main sequence
solar-type stars \citep{duquennoy1991,raghavan2010}, and at even
larger separations for PMS binaries \citep{mathieu1994}. Thus, the
EX~Lup binary with its $\sim$7\,d orbit and $\sim$0.06\,AU separation
may indeed be a very atypical system. Considering companion masses,
\citet{grether2006} found a deficit of companions to Sun-like stars at
around 30-40\,M$_{\rm Jupiter}$, the so-called brown dwarf
desert. Their results show that 11\% of the solar-type stars have
stellar companions and 5\% of them have planetary companions, but less
than 1\% of them have brown dwarf companions. \citet{sahlmann2011} put
an even more stringent upper limit of 0.6\% for the frequency of brown
dwarf companions around Sun-like stars. \citet{endl2006} found that
planets around M dwarfs are even less frequent ($<$1.3\%) than around
FGK-type stars. Depending on the inclination and the exact mass of the
primary, the companion of EX~Lup may fall in the brown dwarf desert,
towards the lower end of the brown dwarf mass range. It is interesting
to speculate about the companion being a deuterium-burning planet
\citep[e.g.,][]{mordasini2009,spiegel2011}.

The existence of a companion around EX~Lup might be related to the
episodic accretion behavior of this star. Close companions are known
to affect the short-term variability of the accretion rate in some PMS
binaries. DQ~Tau and UZ~Tau~E, for example, show signs of pulsed
accretion, a process where the companion periodically modulates the
mass infall from the disk \citep{mathieu1997,jensen2007}. The detailed
analysis of the accretion process in the quiescent EX~Lup system will
be analyzed in a later paper (Sicilia-Aguilar et al.~in prep.). It is
tempting to speculate that the companion may also have a role in the
large outbursts of EX~Lup, like the ones in 1955-1956, and in 2008. It
may slow down the accretion onto the star causing the pile-up of such
a large amount of material that it will eventually result in an
extreme accretion outburst via some kind of instability.

%-----------------------------------------------------------------
% SUMMARY
%-----------------------------------------------------------------

\section{Summary and outlook}

In this paper we presented a five-year RV monitoring study of
EX~Lup. We discovered periodic variations in the RVs of the
photospheric absorption lines. We checked that none of the usual
activity indicators shows periodicity. However, the RV of the narrow
emission lines seem to show slight variations in anti-phase with the
absorption line RVs, if phase-folded with the same period. Based on
our simple modeling, and considering the low $v \sin i <
$3\,km\,s$^{-1}$ and large RV semi-amplitude of 2.2\,km\,s$^{-1}$, we
suggest that a cold or hot spot on the stellar surface is unlikely to
explain the observed absorption line RVs and photometric
variations. The RV signal could be fitted with a companion of $m_2
\sin i$ = 14.7\,M$_{\rm Jup}$ in a 7.417\,d period eccentric orbit
around EX~Lup. Following the analogy with RW\,Aur\,A, we
qualitatively discussed two scenarios to explain the observed spectral
variations in EX~Lup: a geometry with two accretion columns rotating
with the star, and a single accretion flow synchronized with the
orbital motion of the hypothetical companion. Taking 40--50$^{\circ}$
as the most likely value for the inclination, the mass of this
hypothetical companion is probably in the brown dwarf range. The small
separation and large mass ratio makes the EX~Lup a very atypical
binary system. The companion candidate may be responsible for the
smaller or larger accretion outbursts of EX~Lup, supporting those
theories that assume a companion as the triggering mechanism for the
eruptions of certain EXors.

%-----------------------------------------------------------------
% ACKNOWLEDGEMENTS
%-----------------------------------------------------------------

\begin{acknowledgements}
The authors thank the referee, D.~Lorenzetti, for his comments that
helped to improve the manuscript. The authors also thank
V.~Roccatagliata, and M.~Fang for their help with the FEROS
barycentric correction and A.~Simon for his help with the bisector
analysis. This work is based in part on observations made with the
Spitzer Space Telescope, which is operated by the Jet Propulsion
Laboratory, California Institute of Technology, under a contract with
NASA. ASA acknowledges support of the Spanish MICINN/MINECO
``Ram\'{o}n y Cajal'' program, grant number RYC-2010-06164, and the
action ``Proyectos de Investigaci\'{o}n fundamental no orientada'',
grant number AYA2012-35008. This work was partly supported by the
grant OTKA K101393 of the Hungarian Scientific Research Fund.

\end{acknowledgements}

%-----------------------------------------------------------------
% BIBLIOGRAPHY
%-----------------------------------------------------------------
\bibliographystyle{aa}
\bibliography{paper}{}

%-----------------------------------------------------------------
% ONLINE MATERIAL
%-----------------------------------------------------------------

\Online
\vspace*{70mm}
\hspace*{65mm}
\mbox{\LARGE{Online Material}}

\onltab{2}{
\begin{table*}
\caption{FEROS and HARPS observations of EX~Lup}\label{tab:rv}
\begin{tabular}{ccccccccc}
\hline \hline
Date       & Program ID    & PI             & Instrument & MJD             & RV (km\,s$^{-1})$      & $\sigma_{\rm RV}$ (km\,s$^{-1})$ & RV$_{\rm em}$ (km\,s$^{-1})$      & $\sigma_{\rm RV\,em}$ (km\,s$^{-1})$ \\
\hline
2007-07-28 & 079.A-9017(A) & Setiawan       & FEROS      & 54309.115       & $-$1.766               & 0.197                  & $-$1.020 & 0.177 \\
2007-07-29 &               &                &            & 54310.156       & $-$0.732               & 0.151                  & $-$0.063 & 0.251 \\
2007-07-30 &               &                &            & 54311.187       & $-$0.344               & 0.174                  &    1.425 & 0.287 \\
2008-04-21 & 081.A-9005(A) & Zechmeister    & FEROS      & 54577.255       & \dots\tablefootmark{a} & \dots\tablefootmark{a} &  	&	\\
2008-04-21 &	 	   &		    &            & 54577.277       & \dots\tablefootmark{a} & \dots\tablefootmark{a} &  	&	\\
2008-05-05 &               &                &            & 54591.426       & \dots\tablefootmark{a} & \dots\tablefootmark{a} &  	&	\\
2008-05-06 &               &                &            & 54592.152       & \dots\tablefootmark{a} & \dots\tablefootmark{a} &  	&	\\
2008-05-08 & 081.C-0779(A) & Weise          & HARPS      & 54594.360       & \dots\tablefootmark{a} & \dots\tablefootmark{a} &          &       \\
2008-06-16 & 081.A-9023(A) & Weise          & FEROS      & 54633.084       & \dots\tablefootmark{a} & \dots\tablefootmark{a} &  	&	\\
2008-06-23 &               &                &            & 54640.311       & \dots\tablefootmark{a} & \dots\tablefootmark{a} &  	&	\\
2008-07-09 &               &                &            & 54656.977       & \dots\tablefootmark{a} & \dots\tablefootmark{a} &  	&	\\
2009-02-12 & 082.C-0390(A) & Weise          & HARPS      & 54874.338       & $-$0.312               & 0.222                  & $-$0.344 & 0.257 \\
2009-02-13 &               &                &            & 54875.350       & 0.746                  & 0.26                   & $-$2.007 & 0.289 \\
2009-02-14 &               &                &            & 54876.308       & 1.896                  & 0.355                  & $-$1.845 & 0.190 \\
2009-02-15 &               &                &            & 54877.291       & 0.743                  & 0.297                  & $-$1.085 & 0.162 \\
2009-02-15 &               &                &            & 54877.406       & 0.899                  & 0.396                  &  	&	\\
2009-03-01 & 082.C-0427(C) & D\"ollinger    & HARPS      & 54891.300       & 1.605                  & 0.523                  & $-$1.526 & 0.147 \\
2009-03-01 &               &                &            & 54891.382       & 1.908                  & 0.411                  & $-$1.699 & 0.130 \\
2009-03-02 &               &                &            & 54892.274       & 1.205                  & 0.373                  & $-$1.064 & 0.144 \\
2009-03-04 &               &                &            & 54894.260       & $-$3.305               & 0.34                   &    2.230 & 0.283 \\
2009-08-10 & 083.A-9011(B) & Launhardt      & FEROS      & 55053.175       & 0.847                  & 0.142                  & $-$0.290 & 0.444 \\
2009-08-13 &               &                &            & 55056.040       & \dots\tablefootmark{b} & \dots\tablefootmark{b} & \dots\tablefootmark{b} & \dots\tablefootmark{b} \\
2009-08-14 &               &                &            & 55057.993       & $-$2.115               & 0.179                  &    0.862 & 0.519 \\
2009-09-02 & 083.A-9017(B) & Weise	    & FEROS      & 55076.036       & \dots\tablefootmark{b} & \dots\tablefootmark{b} & \dots\tablefootmark{b} & \dots\tablefootmark{b} \\
2010-03-02 & 084.A-9011(B) & Launhardt      & FEROS      & 55257.320       & $-$2.992               & 0.516                  &    0.085 & 0.338 \\
2010-03-04 &               &                &            & 55259.377       & $-$0.745               & 0.115                  &    0.181 & 0.268 \\
2010-03-07 &               &                &            & 55262.307       & 1.535                  & 0.104                  & $-$0.354 & 0.242 \\
2010-03-08 &               &                &            & 55263.368       & $-$0.033               & 0.124                  & $-$0.493 & 0.191 \\
2010-04-23 & 085.A-9027(C) & Gredel         & FEROS      & 55309.394       & $-$2.278               & 0.234                  &    0.247 & 0.199 \\
2010-05-22 &               &                &            & 55338.291       & $-$1.558               & 0.245                  & $-$0.054 & 0.625 \\
2010-06-06 &               &                &            & 55353.327       & $-$1.533               & 0.221                  &    0.135 & 0.224 \\
2010-06-09 &               &                &            & 55356.339       & \dots\tablefootmark{b} & \dots\tablefootmark{b} & \dots\tablefootmark{b} & \dots\tablefootmark{b} \\
2010-06-11 &               &                &            & 55358.271       & \dots\tablefootmark{b} & \dots\tablefootmark{b} & \dots\tablefootmark{b} & \dots\tablefootmark{b} \\
2010-07-19 &               &                &            & 55396.089       & 1.965                  & 0.554                  &    0.104 & 0.177 \\
2010-07-19 &               &                &            & 55396.137       & 1.522                  & 0.587                  &    0.277 & 0.160 \\
2010-07-20 &               &                &            & 55397.111       & $-$0.937               & 0.908                  &    0.812 & 0.148 \\
2010-07-22 &               &                &            & 55399.158       & $-$3.319               & 0.223                  &    1.209 & 0.222 \\
2010-07-23 &               &                &            & 55400.036       & $-$1.888               & 0.152                  & $-$0.789 & 0.186 \\
2010-07-24 &               &                &            & 55401.006       & $-$0.696               & 0.191                  & $-$1.501 & 0.207 \\
2010-07-25 &               &                &            & 55402.084       & 0.297                  & 0.399                  & $-$0.157 & 0.261 \\
2010-07-26 &               &                &            & 55403.112       & 1.586                  & 0.201                  & $-$0.762 & 0.228 \\
2010-07-28 &               &                &            & 55405.161       & $-$3.296               & 2.079                  &    1.057 & 0.196 \\
2010-07-30 &               &                &            & 55407.120       & $-$2.750               & 0.345                  & $-$0.378 & 0.162 \\
\hline
\end{tabular}\\
\tablefoottext{a}{RV was not determined because the source was in outburst.}\\
\tablefoottext{b}{RV was not determined because of low signal-to-noise
  spectrum.}
\end{table*}
}

\onltab{2}{
\begin{table*}
\caption{{\it cont.}}
\begin{tabular}{ccccccccc}
\hline \hline
Date       & Program ID    & PI             & Instrument & MJD             & RV (km\,s$^{-1})$      & $\sigma_{\rm RV}$ (km\,s$^{-1})$ & RV$_{\rm em}$ (km\,s$^{-1})$      & $\sigma_{\rm RV\,em}$ (km\,s$^{-1})$ \\
\hline
2011-01-25 & 086.A-9006(A) & Setiawan       & FEROS      & 55586.371       & 0.297                  & 0.162                  &    0.006 & 0.203 \\
2011-01-26 &               &                &            & 55587.353       & 1.521                  & 0.187                  & $-$0.209 & 0.256 \\
2011-01-27 &               &                &            & 55588.358       & 2.754                  & 1.052                  & $-$0.905 & 0.221 \\
2011-03-10 & 086.A-9012(A) & Setiawan       & FEROS      & 55630.316       & $-$0.210               & 0.123                  & $-$1.444 & 0.294 \\
2011-03-14 &               &                &            & 55634.360       & \dots\tablefootmark{b} & \dots\tablefootmark{b} & \dots\tablefootmark{b} & \dots\tablefootmark{b} \\
2011-03-17 &               &                &            & 55637.395       & $-$1.016               & 0.104                  & $-$1.259 & 0.184 \\
2011-04-16 & 087.A-9013(A) & Mo\'or         & FEROS      & 55667.201       & $-$1.110               & 0.115                  &        &       \\
2011-04-17 &               &                &            & 55668.191       & 0.114                  & 0.123                  & $-$0.632 & 0.213 \\
2011-04-18 &               &                &            & 55669.237       & 1.490                  & 0.266                  &    0.658 & 0.133 \\
2011-04-19 &               &                &            & 55670.294       & 1.951                  & 0.772                  &    0.507 & 0.141 \\
2011-04-20 & 087.A-9029(A) & Gredel         & FEROS      & 55671.198       & $-$0.292               & 1.44                   &    0.932 & 0.132 \\
2011-04-21 &               &                &            & 55672.376       & $-$3.900               & 0.355                  &    2.232 & 0.154 \\
2011-04-22 &               &                &            & 55673.349       & $-$2.974               & 0.155                  &    0.916 & 0.242 \\
2011-04-23 &               &                &            & 55674.324       & $-$1.759               & 0.078                  & $-$0.096 & 0.346 \\
2012-07-03 & 089.A-9007(A) & Mohler         & FEROS      & 56111.013       & $-$2.265               & 0.102                  &    4.46  & 2.2	\\
2012-07-05 & 		   & 		    &	         & 56113.065	   & $-$0.247	   	    & 0.168 	             & $-$0.809 & 0.356 \\
2012-07-05 & 		   & 		    &	         & 56113.977	   & 0.652	   	    & 0.170 	             & $-$0.004 & 0.646 \\
2012-07-06 & 		   & 		    &	         & 56114.977	   & 1.667	   	    & 0.154 	             & $-$1.819 & 0.299 \\
2012-07-08 & 		   & 		    &	         & 56116.024	   & 1.024	   	    & 0.209 	             & $-$0.545 & 0.292 \\
2012-07-09 & 		   & 		    &	         & 56117.062	   & $-$2.144	   	    & 0.157 	             & $-$0.017 & 0.243 \\
2012-07-10 & 		   & 		    &	         & 56118.170	   & $-$2.091	   	    & 0.900 	             &    0.317 & 0.242 \\
2012-07-10 & 		   & 		    &	         & 56118.994	   & $-$1.767	   	    & 0.776 	             &    0.113 & 0.809 \\
2012-07-12 & 		   & 		    &	         & 56120.008	   & $-$0.936	   	    & 0.093 	             & $-$0.377 & 0.580 \\
2012-07-13 & 		   & 		    &	         & 56121.079	   & 0.117	   	    & 0.101 	             & $-$1.292 & 0.764 \\
\hline
\end{tabular}\\
\tablefoottext{a}{RV was not determined because the source was in outburst.}\\
\tablefoottext{b}{RV was not determined because of low signal-to-noise
  spectrum.}
\end{table*}
}

\onltab{3}{
\begin{table*}
\caption{Optical and infrared photometry of EX~Lup in magnitudes}\label{tab:photo}
\begin{tabular}{cccccccc}
\hline \hline
Date       & MJD      & $V$       & $J$      & $H$     & $K^{\prime}$ & [3.6]    & [4.5]    \\
\hline

2010-04-24 & 55310.62 &           &          &         &              & 8.033(7) & 7.590(4) \\
2010-04-25 & 55311.25 & 12.97(5)  & 9.70(2)  & 8.98(2) & 8.56(2)      &          &	   \\ 
2010-04-25 & 55311.28 &           &          &         &              & 8.007(8) & 7.565(3) \\
2010-04-26 & 55312.26 &           &          &         &              & 7.973(2) & 7.525(5) \\
2010-04-26 & 55312.31 & 12.9294)  & 9.71(3)  & 8.96(2) & 8.56(1)      &          &	   \\ 
2010-04-27 & 55313.37 &           &          &         &              & 7.988(3) & 7.516(5) \\
2010-04-27 & 55313.41 & 13.05(4)  &          &         &              &          &	   \\
2010-04-28 & 55314.15 &           &          &         &              & 8.016(13)& 7.532(8) \\
2010-04-29 & 55315.00 &           & 9.69(12) & 9.06(4) & 8.64(4)      &          &	   \\ 
2010-04-29 & 55315.37 &           &          &         &              & 8.059(3) & 7.562(9) \\
2010-04-30 & 55316.21 &           & 9.82(8)  & 9.07(2) & 8.66(2)      &          &	   \\ 
2010-04-30 & 55316.57 &           &          &         &              & 8.021(3) & 7.533(6) \\
2010-05-01 & 55317.19 & 13.22(10) & 9.79(3)  & 8.99(5) & 8.57(1)      &          &	   \\ 
2010-05-01 & 55317.71 &           &          &         &              & 7.969(9) & 7.466(7) \\
2010-05-02 & 55318.80 &           &          &         &              & 7.927(2) & 7.434(4) \\
2010-05-03 & 55319.24 & 13.00(4)  & 9.75(1)  & 9.00(1) & 8.50(4)      &          &	   \\ 
2010-05-03 & 55319.45 &           &          &         &              & 7.871(10)& 7.416(6) \\
2010-05-04 & 55320.24 & 13.02(10) & 9.67(3)  & 8.95(3) & 8.52(2)      &          &	   \\ 
2010-05-04 & 55320.54 &           &          &         &              & 7.833(1) & 7.360(5) \\
2010-05-05 & 55321.25 &           & 9.76(2)  & 9.01(2) & 8.54(5)      &          &	   \\ 
2010-05-05 & 55321.77 &           &          &         &              & 7.841(7) & 7.340(1) \\
2010-05-06 & 55322.40 & 13.25(2)  & 9.85(3)  & 9.08(2) & 8.57(5)      &          &	   \\ 
2010-05-07 & 55323.09 &           &          &         &              & 7.860(10)& 7.382(6) \\
2010-05-07 & 55323.97 &           &          &         &              & 7.899(9) & 7.421(6) \\
2010-05-08 & 55324.22 & 13.21(11) & 9.81(1)  & 9.08(2) & 8.57(1)      &          &	   \\ 
2010-05-09 & 55325.22 & 13.14(5)  & 9.78(5)  & 8.96(3) & 8.60(3)      &          &	   \\ 
2010-05-10 & 55326.22 & 12.97(6)  & 9.72(1)  & 8.94(3) & 8.52(2)      &          &	   \\ 
\hline
\end{tabular}
\end{table*}
}

%-----------------------------------------------------------------
% END
%-----------------------------------------------------------------
\end{document}